\documentclass[twocolumn,showpacs,pre,floatfix,amsmath,amssymb,notitlepage]{revtex4-1}
\usepackage{epsfig}
\usepackage{subfigure}
\usepackage{amsmath}
\usepackage{color}
\usepackage{amssymb}
\usepackage{setspace}
\usepackage{graphicx}
\usepackage{dcolumn}
\usepackage{bm}
\usepackage{times}
\usepackage{enumerate}
\usepackage{footnote}
\input epsf

\graphicspath{{figures/}}

\begin{document}

\author{Linpeng Gu$^{1}$}
\author{Hanlin Fang$^{2}$}
\author{Juntao Li$^{2}$} 
\author{Liang Fang$^{1}$} 
\author{Soo Jin Chua$^{3}$} 
\author{Jianlin Zhao$^{1*}$, Xuetao Gan$^{1}$} 
\email{jlzhao@nwpu.edu.cn; xuetaogan@nwpu.edu.cn} 
\affiliation{$^{1}$MOE Key Laboratory of Material Physics and Chemistry under Extraordinary Conditions, and Shaanxi Key Laboratory of Optical Information Technology, School of Science, Northwestern Polytechnical University, Xi'an 710072, China}
\affiliation{$^{2}$State Key Laboratory of Optoelectronic Materials and Technologies, School of Physics, Sun Yat-Sen University, Guangzhou, 510275, China}
\affiliation{$^{3}$Department of Electrical and Computer Engineering, National University of Singapore, 4 Engineering Drive 3, Singapore 117583}

\date{\today}

\title{A compact structure for realizing Lorentzian, Fano  and EIT resonance lineshapes in a microring resonator}


\maketitle

{\bf	Microring resonators, as a fundamental building block of photonic integrated circuits, have been well developed into numerous functional devices, whose performances are strongly determined by microring's resonance lineshapes. We propose a compact structure to reliably realize Lorentzian, Fano, and electromagnetically induced transparency (EIT) resonance lineshapes in a microring. By simply inserting two air-holes in the side-coupled waveguide of a microring, a Fabry-Perot (FP) resonance is involved to couple with microring's resonant modes, showing Lorentzian, Fano, and EIT lineshapes over one free spectral range of the FP resonance. The quality factors, extinction ratios, and slope rates in different lineshapes are discussed. At microring's specific resonant wavelength, the lineshape could be tuned among these three types by controlling the FP cavity's length. Experiment results verify the theoretical analysis well and represent Fano lineshapes with  extinction ratios of about 20 dB and slope rates over 280 dB/nm. The reliably and flexibly tunable lineshapes in the compact structure have potentials to improve microring-based devices and expand their application scopes.
}\\

\section{Introduction}\label{sec:01}

Microring resonators (MRRs) have been widely employed for on-chip optical interconnects, nonlinear optics, and sensings relying on their compact size, high quality ($Q$) factor, and compatibility with the integration of other passive and active photonic devices~\cite{Lipson2018, Alex2018, sun2015single, kippenberg2011microresonator, ferdous2011spectral, zhang2016highly}. MRRs are normally side-coupled with a bus-waveguide to access their resonant modes, yielding periodic resonant dips (usually symmetric Lorentzian-type) in the bus-waveguide's transmission spectra. To carry out MRR-based optical sensing, filtering, modulation, and switching, transmission variations of the bus-waveguide around the resonant dips are required, which are achieved by shifting MRR's resonance either toward or away from the signal wavelength. To obtain a large transmission variation, a shift of the resonant wavelength larger than the linewidth of the resonant dip is desired~\cite{Fanoreview, chao2003biochemical, fan2002sharp}. Thus, lineshapes of MRR's resonant dips, i.e. sharpness and linewidth, would strongly determine the performances of MRR-based devices, such as power consumption, sensing sensitivity, modulation depth, extinction ratio, etc~\cite{fan2002sharp, mario2006asymmetric}. 

Considerable works have been implemented to modify MRR's resonance lineshape into a Fano-type, which has an asymmetric and sharp slope around the resonant wavelength. As opposed to the usual symmetric Lorentzian resonances, the wavelength range for tuning the Fano resonance transmission from 0\% to 100\% could be much narrower than the full linewidth of the resonance itself~\cite{fan2002sharp}. It therefore enables devices with improved performances, such as all-optical switches with one order of magnitude energy reduction and a ratio-metric wavelength monitor with an ultrahigh resolution of 0.8 pm~\cite{heuck2013improved, mario2006asymmetric, wang2016fano}. Fano resonances in MRRs are normally realized by interfering the resonance pathway with a coherent background pathway. A straightforward approach is the utilization of Mach-Zehnder interferometers (MZIs), where MRRs are either side-coupled to one arm of a MZI or inserted into a MZI to cause the two resonant beams propagating in the MRRs to interfere. MZI's excess waveguide-arms allow easy tunability of Fano lineshapes. Unfortunately, MRR devices incorporating MZI lose their compactness. By coherently coupling multiple resonant modes from multi-MRRs into one bus-waveguide, Fano resonances could also be realized. Nevertheless, it is challenging to achieve precise structure designs and device fabrications~\cite{tobing2008box, darmawan2005phase}. Also, more MRRs generate a larger footprint. 

It is also possible to alter the MRR's resonance into an EIT lineshape, showing a narrow transparency peak residing in a broad transmission valley. It has potentials in ultra-dense on-chip wavelength division multiplexer~\cite{mancinelli2011coupled} as well as enhancing the cavity's finesse~\cite{tobing2008finesse}. EIT resonances originate from coherent interferences between coupled resonant modes. They are realized in structures consisting of two or more MRRs coupled with a bus-waveguide or a MZI~\cite{darmawan2008resonance, mancinelli2011coupled, smith2004coupled, totsuka2007slow}, which have large device footprint and require rigorous considerations of the resonant modes' overlap. {In addition, while Lorentzian, Fano, and EIT resonance lineshapes have different functionalities, to the best of our knowledge, a MRR-based structure supporting these three lineshapes simultaneously has not been reported. Also, the modification of resonance lineshapes at a specific wavelength among the three types by simply tuning the structure parameters is desired.}

\begin{figure}[htbp]
	\centering
	\includegraphics[width=\linewidth]{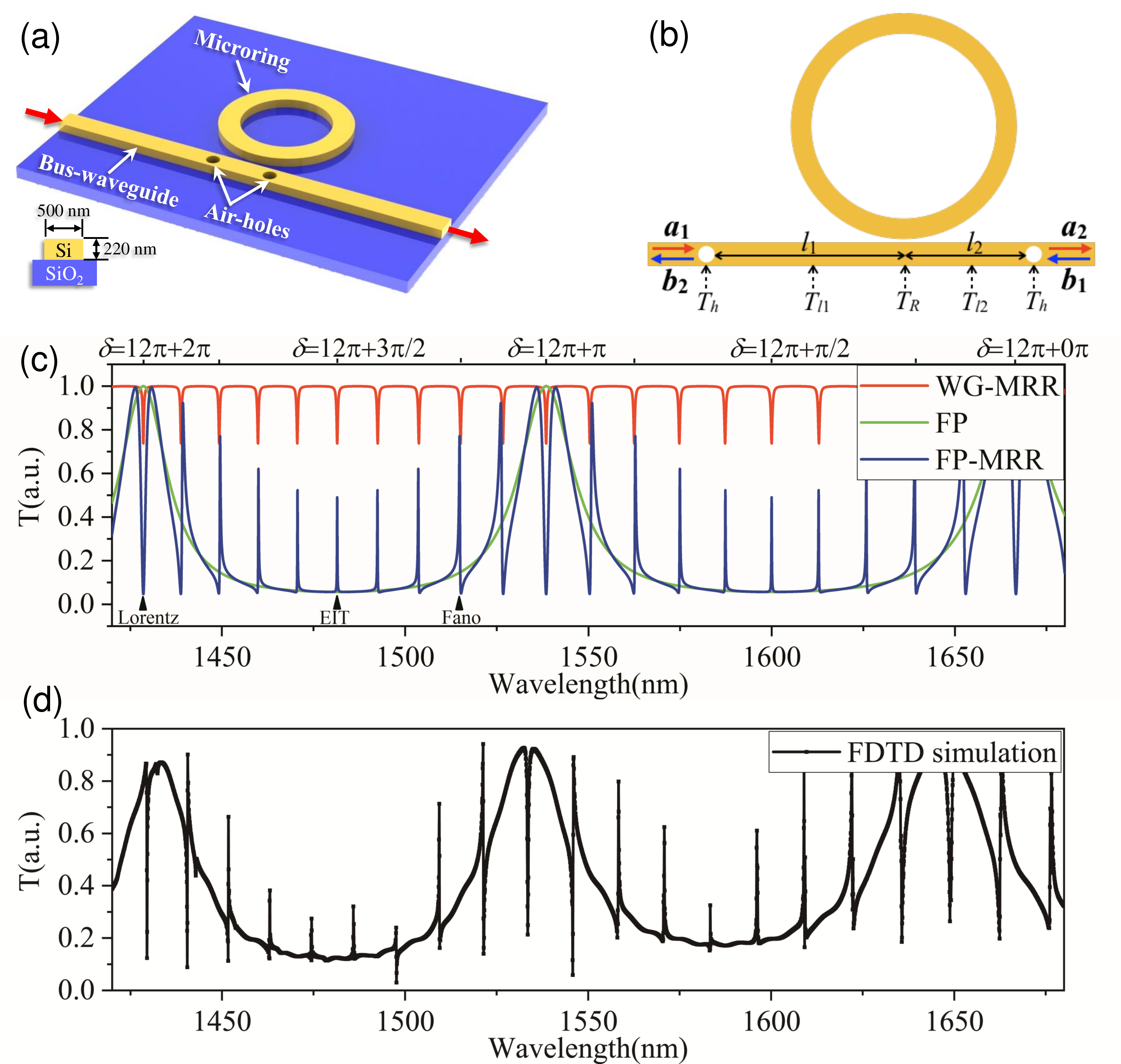}
	\caption{ (a) Schematic and (b) theoretical model of the proposed structure, where the MRR is coupled with a bus-waveguide inserted with two air-holes. (c) Calculated transmission spectra of a bus-waveguide with two air-holes (FP), and a MRR coupled with a bus-waveguide with (FP-MRR) and without (WG-MRR) air-holes. {(d) Simulated transmission spectrum of the FP-MRR structure using a FDTD method.} }
	\label{fig:model}
\end{figure}

In this letter, we demonstrate that, by simply coupling a MRR with a bus-waveguide inserted with two air-holes, as shown  schematically in Fig.~\ref{fig:model}(a), all of the above mentioned resonance lineshapes could be realized. The structure is compact and promises great design and fabrication tolerances. The two air-holes constitute a low-finesse FP cavity  in the bus-waveguide, producing broadband resonant peaks to couple with MRR's multiple resonant modes. {Here, the air-holes are designed with circular shapes, considering their less fabrication imperfections than air-holes with triangular or rectangular shapes.} The transmission spectrum of the coupled system is analyzed by the transfer matrix method. Lorentzian, Fano, and EIT resonance lineshapes are obtained over a free spectral range (FSR) of the FP cavity. For a specific resonant mode of the MRR, its lineshape could be tuned readily among the three types by changing the distance between the two air-holes. The results of the theoretical analysis are verified experimentally by devices fabricated on a silicon-on-insulator chip. Different from the previously reported structures, the proposed design has only one MRR and a bus-waveguide, and thus making it very compact. In addition, the  broadband resonances of the FP cavity facilitate their overlaps with MRR's narrowband resonances. Thence, it is not necessary to carefully consider the design and fabrication of the two air-holes.

\section{Model and theory}\label{sec:02}

The transmission spectrum of the proposed structure could be obtained by applying a transfer matrix analysis on the model shown in Fig.~\ref{fig:model}(b)~\cite{fan2002sharp}. The bus-waveguide has the forward (backward) propagating mode with the input and output of $a_1$ ($b_1$) and $a_2$ ($b_2$), respectively. The two air-holes work as partial reflectors of the propagating modes in the bus-waveguide. For  simplicity, they are designed with the same diameters. With an amplitude reflection coefficient of $r_h$, one of the air-holes generates a transfer matrix for the propagating mode of
\begin{equation}
M_h=\frac{1}{i\sqrt{1-{r_h}^2}}\begin{bmatrix}
{-1}&{-r_h}\\
{r_h}&1
\end{bmatrix}
\end{equation}
The two air-holes are located at distances of $l_1$ and $l_2$ to the waveguide-MRR coupling point. When light propagates through the two waveguide sections, the transfer matrixes are determined by the phase shifts in the form of 
\begin{equation}
M_{l_1}=\begin{bmatrix}
{e^{i2\pi{nl_1}/\lambda}}&{0}\\
{0}&{e^{-i2\pi{nl_1}/\lambda}}
\end{bmatrix}, 
M_{l_2}=\begin{bmatrix}
{e^{i2\pi{nl_2}/\lambda}}&{0}\\
{0}&{e^{-i2\pi{nl_2}/\lambda}}
\end{bmatrix} 
\end{equation}
Here, $n$ is the effective refractive index of the propagating mode, and $\lambda$ is the operating wavelength. In MRRs, the resonances are traveling-waves, and the backward scattering in the waveguide-MRR coupling could be neglected~\cite{bogaerts2012silicon}. For a forward or backward waveguide mode propagating through the coupling point, its transmission spectrum is governed by $t_R(\lambda)=\frac{t-ae^{i2\pi{nL_R}/\lambda}}{1-tae^{i2\pi{nL_R}/\lambda}}$~\cite{heebner2008optical}, where  $t$ is the field transmission coefficient at the waveguide-MRR coupling region, and $a=\exp(-\alpha{L_R})$ is the round trip amplitude for a MRR with perimeter $L_R$ and linear loss coefficient $\alpha$. The corresponding transfer matrix for an incident light after the directional waveguide-MRR coupling could be described as 
\begin{equation}
M_R=\begin{bmatrix}
{t_R}&{0}\\
{0}&1
\end{bmatrix}
\end{equation}
The transfer matrix equation for the incoming and outgoing wave amplitudes of the entire coupled system is governed by 
\begin{equation}
\begin{bmatrix}
b_1\\
a_2\end{bmatrix}=M_hM_{l_2}M_RM_{l_1}M_h\begin{bmatrix}
a_1\\
b_2 \end{bmatrix}
\end{equation}
To be consistent with the common operations of  MRR-based devices, we consider only the left input port has an incoming normalized light, i.e., $a_1=1$, $b_1=0$. The final power transmission spectrum of the coupled system could be calculated as 
\begin{equation}
T(\lambda)=\left|{\frac{a_2}{a_1}}\right|^2=\left|\frac{{(1-r^2_h)}{t_R(\lambda)}e^{\frac{i2\pi{nl}}\lambda}}{1-{r^2_h}{t_R(\lambda)}e^{\frac{i4\pi{nl}}\lambda}}\right|^2
\label{eq:T}
\end{equation}
where $l=l_1+l_2$. 

We could transform Eq.~\ref{eq:T} into a straightforward form to indicate the coupling between the MRR and FP cavity, as shown below 
\begin{equation}
T(\lambda)=\left|\frac{{(1-r^2_h)}e^{\frac{i2\pi{nl}}\lambda}}{1-{r^2_h}e^{\frac{i4\pi{nl}}\lambda}}\right|^2\left|t_R(\lambda)\right|^2\left|\frac{1-{r^2_h}e^{\frac{i4\pi{nl}}\lambda}}{1-{r^2_h}{t_R(\lambda)}e^{\frac{i4\pi{nl}}\lambda}}\right|^2
\label{eq:T6}
\end{equation}

On the right side of Eq.~\ref{eq:T6}, the first and second terms are sole transmissions of the FP cavity and the side-coupled MRR, respectively; the third term indicates their resonance coupling. By assuming $r_h=0$,  i.e., no air-holes in the bus-waveguide, the transmission spectrum of a conventional bus-waveguide coupled MRR could be calculated from Eq.~\ref{eq:T6}, as displayed in the red curve (WG-MRR) of Fig.~\ref{fig:model}(c). Here, the parameters of the waveguide-coupled MRR are $t=0.9$, $a=0.99$, $nL_R=200~\mu$m, and the spectrum is calculated over a range of 1,400-1,700 nm. Periodic resonant dips with a FSR of $\sim$12 nm are obtained. Because $t\neq{a}$, the extinction ratio (ER) of the dips is not high, which is calculated as $10\log_{10}({\frac{1-ta}{t-a}})^2=1.67$~dB. Their quality ($Q$) factors determined by $t$ and $a$ are estimated as 3,500. 

In another case, if we modify $t=1$, $a=0$ and set the two air-holes with parameters of $nl=10~\mu$m and $r_h=0.78$, no MRR is coupled with the bus-waveguide. The coupled system's transmission is determined only by the FP cavity of the two air-holes. The calculated transmission from  Eq.~\ref{eq:T6} is plotted as shown by the green curve (FP) in Fig.~\ref{fig:model}(c). When the phase delay between the two air-holes $\delta=2\pi{nl/\lambda}$ is an integer multiple of $\pi$, there are resonant peaks with broad linewidths ($Q$$\approx$87) due to the low reflection coefficients of the air-holes. The ER of the resonant peaks is  $10\log_{10}({\frac{1+r^2_h}{1-r^2_h}})^2=12.47$~dB.

The blue curve in Fig.~\ref{fig:model}(c) displays the transmission spectrum of a coupled system (FP-MRR) with parameters of $t=0.9$, $a=0.99$, $nL_R=200~\mu$m, $nl=10~\mu$m and $r_h=0.78$, which includes both the MRR and FP cavity. Because of couplings between their resonances, Lorentzian, Fano, and EIT resonance lineshapes are realized simultaneously over each FSR of the broad FP oscillation background. At a FP cavity's resonant peak, i.e., $\delta$ is an integer multiple of $\pi$, if there is a resonance of MRR, their coherent overlap gives rise to a symmetric Lorentzian resonant dip. As indicated in Fig.~\ref{fig:model}(c), the Lorentzian resonant dips in the transmissions of WG-MRR and FP-MRR have significantly different ERs and $Q$ factors. While the maximum transmissions around the dips of WG-MRR and FP-MRR are similar, the bottom of the FP-MRR dip reduces to an ultralow value of $0.0443$, resulting in a greatly improved ER of  $10\log_{10}({\frac{1-ta-r^2_h(t-a)}{(1-r^2_h)(t-a)}})^2=13.54$~dB. In addition, for the optical field of the resonant mode in the FP cavity, it will couple with the MRR many times during its backward and forward reflections, which therefore introduces extra waveguide-coupling losses to the MRR and represents a decreased $Q$ factor. The $Q$ factor of  the Lorentzian dip in the FP-MRR is estimated to be 1,000 by a Lorentzian fitting. 

At a specific wavelength, if the phase delay between the two air-holes $\delta$ is an odd multiple of $\pi/2$, the FP cavity is in a completely destructive interference state. A MRR's resonance coupled with this state would generate an EIT transparency peak over the broad transmission valley, as illustrated in Fig.~\ref{fig:model}(c). Calculated from Eq.~\ref{eq:T6}, ER of EIT lineshape has a value of $10\log_{10}({\frac{(1+r^2_h)(t-a)}{1-ta+r^2_h(t-a)}})^2=8.67$~dB, which is larger than the ER of the Lorentzian resonant dips in WG-MRR. Its $Q$ factor is estimated as 13,000 by a Lorentzian fitting, which is much higher than that of the resonant dips in WG-MRR. This increased $Q$ factor in the EIT lineshape represents an improvement of photon's lifetime and finesse of the MRR by adding a FP cavity in the bus-waveguide. An optimized parameter could be derived for a much higher ER; for instance, ER $=$~13 dB is obtained for $r_h=0.9$. The destructive interference between the two air-holes results in a weakened waveguiding mode around the waveguide-MRR coupling region, which therefore reduces the waveguide-MRR coupling coefficient. For light resonant in the MRR, less waveguide-coupling loss therefore promises an improved $Q$ factor. In conventional waveguide-MRR coupling systems, to achieve transmission peaks, an additional drop-waveguide is required to couple with the MRR~\cite{bogaerts2012silicon}, which makes the device complicated and produces more coupling losses in the MRR. In the proposed FP-MRR, the EIT lineshape provides a transmission peak with only one bus-waveguide and the linewidth is much narrower, potentially expanding compact MRR's applications in nonlinear optics, sensing, filtering, and switching. 

\begin{figure}[htbp]
	\centering
	\includegraphics[width=\linewidth]{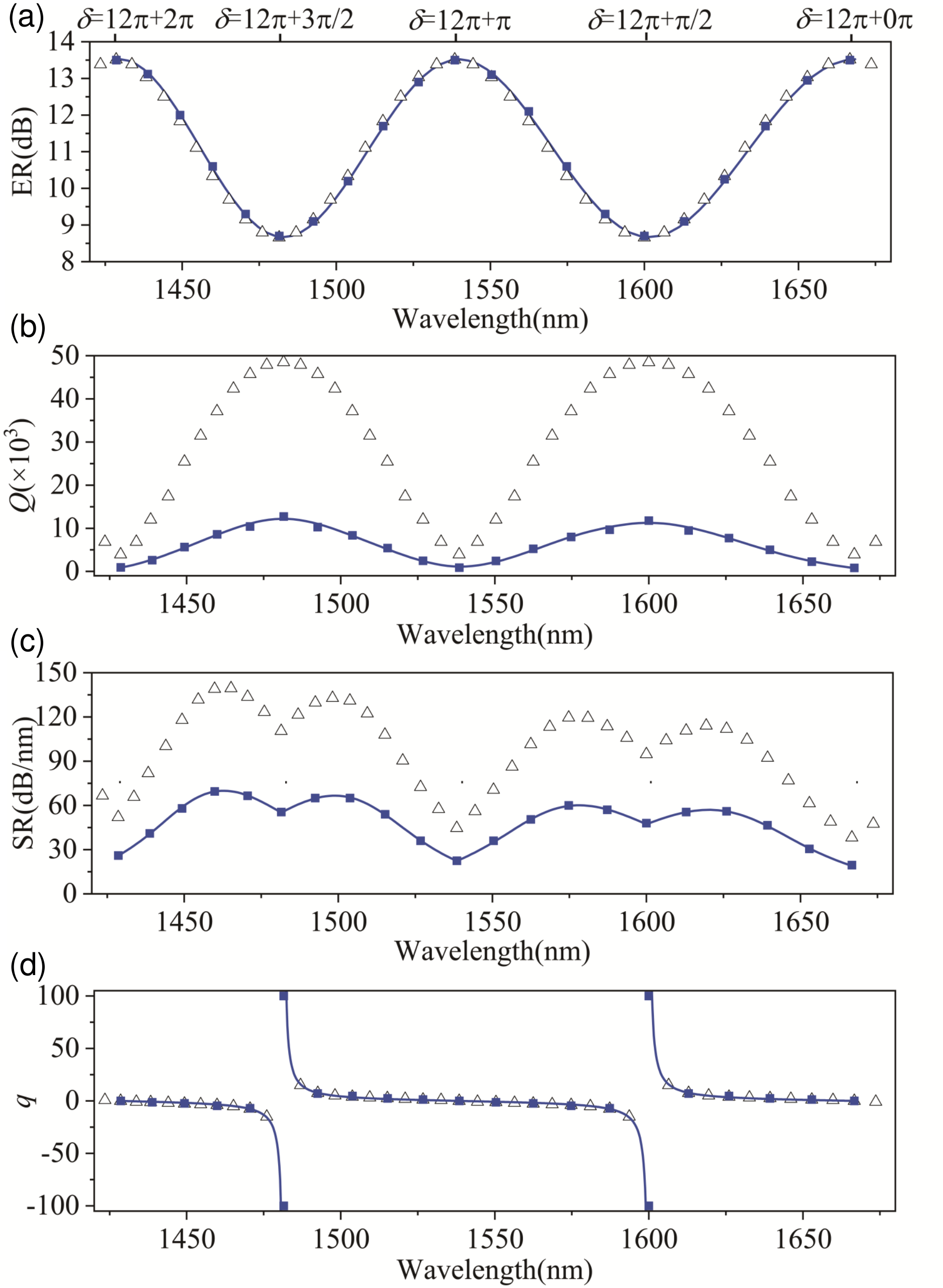}
	\caption{ {(a) ERs, (b) $Q$ factors, (c) SRs, and (d) Fano parameters ($q$) of the different resonance lineshapes in the FP-MRR spectrum with the parameter $nL_R$ equalling to $200~\mu$m (blue curves with squares) and $400~\mu$m (scattered triangular points). } }
	\label{fig:ER}
\end{figure}

When $\delta$ is in the ranges of $(k\pi,k\pi+\pi/2)$ and $(k\pi+\pi/2,k\pi+\pi)$, where $k$ is an integer, the waveguiding modes are off-resonance from the FP cavity. Their couplings with MRR's resonances give rise to Fano resonance lineshapes over the slopes of the FP oscillation background, as shown in Fig.~\ref{fig:model}(c). The Fano resonance lineshapes depend critically on the resonant wavelengths with respect to the FP resonant peaks. At different sides of the FP resonant peaks, the asymmetric directions of the Fano resonances are opposite. 

{The above obtained resonance lineshapes are further verified by mode simulations using a finite difference time domain (FDTD) technique (Lumerical Inc.). In the modelled FP-MRR, the bus-waveguide and MRR have the same height and width of 220 nm and 500 nm, and the refractive index is chosen as 3.4. The MRR has a radius of 7.2 $\mu$m, and is coupled with the bus-waveguide with a gap of 100 nm. Two circular air-holes with a radius of 160 nm are inserted in the bus-waveguide with a distance of 2.8 $\mu$m. The simulated transmission spectrum of the proposed FP-MRR is plotted in Fig.~\ref{fig:model}(d), showing consistent results with the theoretical prediction (blue curve in Fig.~\ref{fig:model}(c)). Lorentzian dips and EIT transparency peaks are obtained at the resonant peaks and transmission valleys of the FP oscillation background, respectively. There are Fano resonance lineshapes over the slopes of FP cavity background.}

To describe distinct features of the three types of lineshapes, we plot ERs, $Q$ factors, slope rates (SRs), and Fano parameters $q$~\cite{Fanoreview} of the resonance lineshapes in the FP-MRR spectrum {(blue curve in Fig.~\ref{fig:model}(c)). }While ERs of the Lorentzian and EIT resonance lineshapes could be deduced  from Eq.~\ref{eq:T6}, Fano resonance's ER has no explicit expression. From the transmission spectrum of the FP-MRR, ERs are calculated numerically by counting the maximum and minimum around each resonance, as shown in Fig.~\ref{fig:ER}(a).  Fano resonances' ERs are between those of the Lorentzian and EIT resonances; ERs for all the resonance lineshapes present a sine-like function with wavelength. $Q$ factors of the three resonance lineshapes are extracted as well by fitting them with Lorentzian or Fano functions, as illustrated in Fig.~\ref{fig:ER}(b). The maximum and minimum $Q$ factors are achieved in EIT and Lorentzian resonances, respectively, while Fano resonances have $Q$ factors lying between them. Different $Q$ factors are attributed to the varied coupling strengths between MRR and waveguiding modes, which are determined by the interference states of the optical field confined by the two air-holes in the bus-waveguide.

SR, as another important figure of merit for resonance lineshapes, is calculated by taking derivations of the slopes in the resonance lineshapes. It determines the wavelength range for tuning the transmission between maximum and minimum. A calculated SR spectrum for the resonance lineshapes is plotted in Fig.~\ref{fig:ER}(c). For the symmetric Lorentzian and EIT resonance lineshapes, the SR is determined by the $Q$ factors, i.e., linewidths. Lorentzian (EIT) resonance with the lower (higher) $Q$ factor has smaller (larger) SR. Note that for Fano resonances with asymmetric lineshapes, the SRs could be higher than that of the EIT resonance, though the $Q$ factors might be lower. It would promise devices with higher performances relying on the much narrower wavelength range for switching the maximum and minimum transmissions, while its ER is not the highest (as shown in Fig.~\ref{fig:ER}(a)). Fano parameter $q$ characterizes the specific asymmetric profile of the Fano lineshape~\cite{Fanoreview}, which is estimated by fitting them with Fano functions, as plotted in Fig.~\ref{fig:ER}(d). A cotangent-like function is obtained for different $q$ at the resonances, showing $q=0$ ($q=\pm{\infty}$) at the Lorentzian (EIT) resonance.  Additionally, both of the Lorentzian dips and EIT peaks in the FP-MRR have the same central wavelengths as those of the Lorentzian dips in the WG-MRR. However, the maximum and minimum of the Fano lineshapes shift  slightly from the WG-MRR's resonant wavelengths due to the complex interference mechanism. 

{We also study the resonance lineshapes of another FP-MRR by changing the MRR's perimeter into $nL_R=400~\mu$m. With the same principle of mode interferences, Lorenzian, Fano,  and EIT resonance lineshapes are obtained. The figure of merits of the resonance lineshapes are displayed in the scattered triangular points of Fig.~\ref{fig:ER}, which have the same trends as those in FP-MRR with perimeter of $nL_R=200~\mu$m (see blue curves with squares in Fig.~\ref{fig:ER}). The calculated ER, $Q$, SR, and $q$ of this FP-MRR with larger perimeter could be explained according to the above analysis. Because of less scattering loss for the larger MRR, the $Q$ factors are increased for the resonant modes. As a result, SRs, which are mainly determined by the resonance linewidth, have much higher values for the FP-MRR with larger MRR. Since the coupling gap between the MRR and bus-waveguide is not changed, the factors of ER and $q$ are almost unmodified.}

\section{Experimental results}\label{sec:03}

\begin{figure}[htbp]
	\centering
	\includegraphics[width=\linewidth]{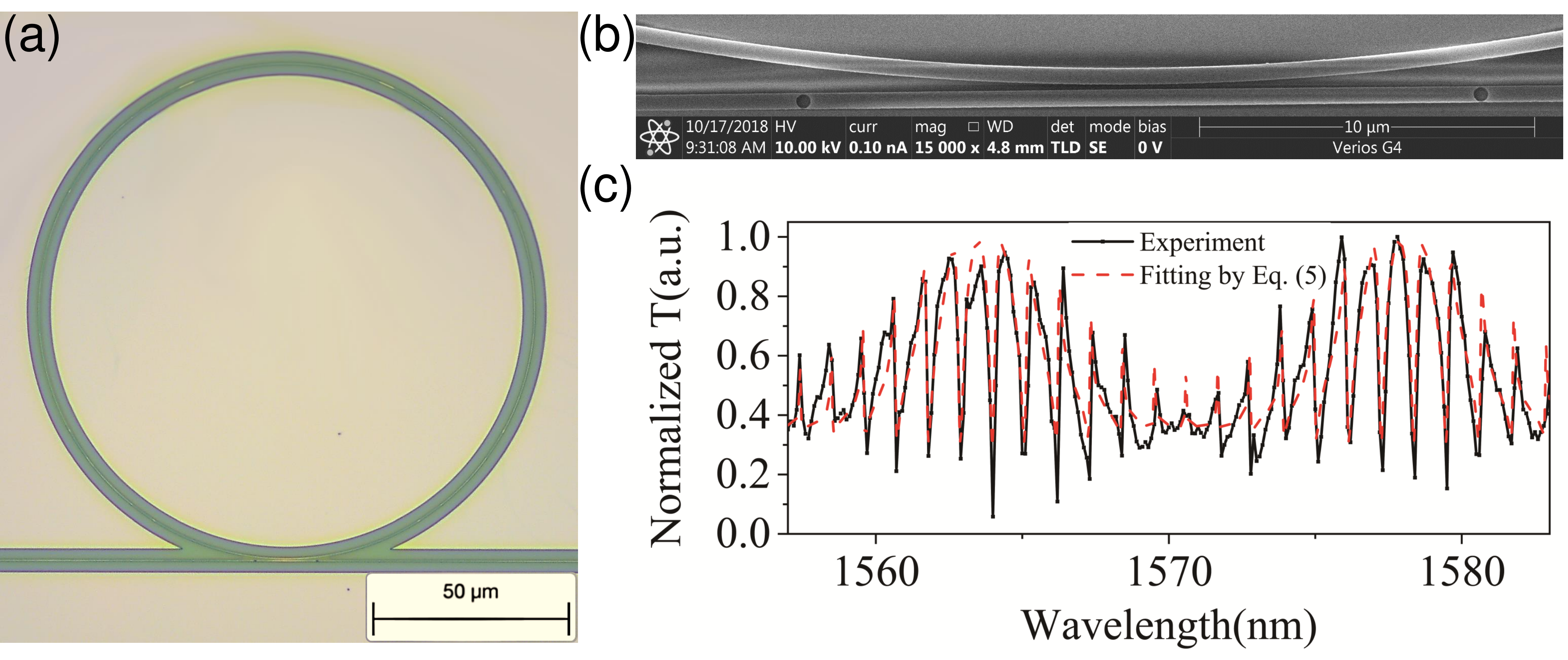}
	\caption{ (a) Optical microscope image of one of the fabricated device; (b) SEM image of the waveguide-MRR coupling region with two air-holes inserted in the bus-waveguide; {(c) Measured transmission spectrum of the fabricated device (black line) and fitting result by Eq.~\ref{eq:T} (red dashed line), showing Lorentzian, Fano, and EIT resonance lineshapes over the FP oscillations. } }
	\label{fig:exp}
\end{figure}

To verify the above theoretical analysis, we fabricate the proposed MRR device on a silicon-on-insulator chip with a 220 nm thick top silicon layer and a 2 $\mu$m thick buried oxide layer. Electron beam lithography is used to define the device patterns, which are then transferred onto the top silicon layer by inductively coupled plasma etching. Grating couplers with two-dimensional air-hole arrays are designed at both ends of the bus-waveguides~\cite{liu2010high}. The bus-waveguide and MRR are designed with the same width of 500 nm, and their coupling gap is 120 nm. Figure~\ref{fig:exp}(a) displays an optical microscope image of the fabricated device, where the MRR has a radius of 80 $\mu$m. The scanning electron microscope (SEM) image of the waveguide-MRR coupling region is shown in Fig.~\ref{fig:exp}(b), displaying the two air-holes with a distance of 20 $\mu$m and a radius of 150 nm. 

The fabricated devices are characterized by coupling a narrowband tunable laser (TL) into the input grating coupler, and the transmission powers are monitored using a photodiode. By tuning the TL wavelength over a range of { 1,557-1,583 nm }in steps of 0.02 nm, the device transmission spectrum could be analyzed. {The black line in Fig.~\ref{fig:exp}(c) }shows the result of the measurement from the device shown in Fig.~\ref{fig:exp}(a). Confined by the two air-holes, the waveguide transmission presents FP oscillations with FSRs of about 14 nm. Superimposed on them are the various lineshapes of the MRR's resonances. {On the top of the FP oscillation(at the wavelengths of 1563.9 nm and 1578.4 nm), the symmetric Lorentzian resonance dips  are observed.} Fano resonance lineshapes with opposite asymmetric directions are formed on both sides of the FP slopes. {At the valley of the FP background (at 1570.5 nm), an EIT transparency peak presents at the transmission valley of the FP oscillation background.} These experimental results are consistent with the predictions in Fig.~\ref{fig:model}.

{As shown in the red dashed line of Fig.~\ref{fig:exp}(c), the experimental result is fitted by the theoretical prediction of Eq.~\ref{eq:T} with parameters of $t=0.75$, $a=0.96$ and $r_h=0.61$, presenting their good consistency. Both of them have Lorentzian, Fano, and EIT resonance lineshapes at the corresponding resonant wavelengths. However, because of fabrication imperfections, ERs of the FP oscillations are lower than the theoretical values since the lower reflectivities of the air-holes, which also causes that the EIT peaks are not as prominent as the theoretical fitting. From the experiment measurements, the maximum ERs and SRs of the Fano resonances are estimated as 20 dB and 280 dB/nm, respectively, averaged over tens of devices. The values calculated from the corresponding theoretical predictions are 30 dB and 450 dB/nm, respectively. By optimizing the fabrication processes in future devices, better figure of merits are expected from the proposed structure. }

\section{Conclusions}\label{sec:04}

In conclusion, we demonstrated a compact structure for realizing Lorentzian, Fano and EIT resonance lineshapes in a waveguide side-coupled MRR. By adding two air-holes into the bus-waveguide, all of the three lineshapes could be obtained simultaneously in the waveguide transimission. For a specific resonance, its  lineshape could be designed reliably by changing the distance between the two air-holes. The different resonance lineshapes achieved flexibly in the proposed structure promise great potentials to expand MRRs' applications in optical interconnects, nonlinear optics, and sensing. 
\\
\section*{Acknowledgments}\label{sec:05}

Financial support was provided by National Natural Science Foundations of China (61522507, 61775183, 11634010); the Key Research and Development Program (2017YFA0303800, 2018YFA0307200); the Key Research and Development Program in Shaanxi Province of China (2017KJXX-12); the Fundamental Research Funds for the Central Universities (3102018jcc034, 3102017jc01001).

\bibliographystyle{plain}

\end{document}